\documentclass[9pt,sigconf,letterpaper]{acmart}

\AtBeginDocument{%
  }

\copyrightyear{2026}
\acmYear{2026}
\setcopyright{cc}
\setcctype{by}
\acmConference[IMC '26]{Proceedings of the 2026 ACM Internet Measurement Conference}{October 12--16, 2026}{Karlsruhe, Germany}
\acmBooktitle{Proceedings of the 2026 ACM Internet Measurement Conference (IMC '26), October 12--16, 2026, Karlsruhe, Germany}
\acmDOI{10.1145/3777912.3809138}
\acmISBN{979-8-4007-2327-8/2026/10}

\usepackage[english]{babel}
\usepackage{blindtext}
\usepackage{subcaption}
\usepackage{tabularx}
\usepackage{tikz}
\usepackage{listings}

\newcommand{\one}{({\em i}\/)\xspace}
\newcommand{\two}{({\em ii}\/)\xspace}
\newcommand{\three}{({\em iii}\/)\xspace}

\newcommand{\pb}[1]{\vspace{0.75ex}\noindent{\bf \em #1}\hspace*{.3em}}
\newcommand*\circled[1]{\tikz[baseline=(char.base)]{
            \node[fill=lightgray!20,shape=circle,draw,inner sep=2pt] (char) {\textbf #1};}}

\begin{document}

\title[NAT Traversal Measurement Campaign]{Large-Scale Measurement of NAT Traversal for the Decentralized Web: A Case Study of DCUtR in IPFS}

\author{Dennis Trautwein}
\orcid{0000-0002-8567-2353}
\affiliation{%
   \institution{University of Göttingen}
  \country{Germany}
}
\email{research@dtrautwein.eu}

\author{Cornelius Ihle}
\orcid{0000-0002-3994-5218}
\affiliation{%
  \institution{University of Göttingen}
  \country{Germany}
}
\email{ihle@gipplab.org}

\author{Moritz Schubotz}
\orcid{0000-0001-7141-4997}
\affiliation{%
  \institution{FIZ Karlsruhe -- Leibniz Institute for Information Infrastructure}
  \country{Germany}
}
\email{moritz.schubotz@fiz-karlsruhe.de}

\author{Corinna Breitinger}
\orcid{0000-0001-6586-0392}
\affiliation{%
  \institution{University of Göttingen}
  \country{Germany}
}
\email{breitinger@uni-goettingen.de}

\author{Bela Gipp}
\orcid{0000-0001-6522-3019}
\affiliation{%
  \institution{University of Göttingen}
  \country{Germany}
}
\email{gipp@uni-goettingen.de}

\renewcommand{\shortauthors}{Trautwein et al.}

\begin{abstract}
    The promise of decentralized peer-to-peer (P2P) systems is fundamentally gated by the challenge of Network Address Translation (NAT) traversal, with existing solutions often reintroducing the very centralization they seek to avoid. This paper presents the first large-scale measurement study of a fully decentralized NAT traversal protocol, Direct Connection Upgrade through Relay (DCUtR), within the production libp2p-based InterPlanetary File System (IPFS) network. Drawing on over 4.4 million traversal attempts from 85,000+ distinct networks across 167 countries, we provide an empirical analysis of modern P2P connectivity. We establish a conditional success rate of $70\% \pm 7.1\%$ for the hole-punching stage, given that prerequisite relay reservation and public address discovery succeed, providing a crucial new benchmark for the field. Critically, we empirically challenge the long-held belief of UDP's superiority for NAT traversal, demonstrating that DCUtR's high-precision, RTT-based synchronization yields statistically indistinguishable success rates for both TCP and QUIC ($\sim70\%$). Our analysis further validates the protocol's design for permissionless environments by showing that success is independent of relay characteristics and that the mechanism is highly efficient, with $97.6\%$ of successful connections established on the first attempt. Building on this analysis, we propose a concrete roadmap of protocol enhancements aimed at achieving universal connectivity and contribute our complete dataset to foster further research in this domain.
\end{abstract}

\ccsdesc[500]{Networks~Protocol testing and verification}
\ccsdesc[500]{Networks~Network protocol design}
\ccsdesc[500]{Networks~Network measurement}
\ccsdesc[500]{Networks~Network experimentation}
\ccsdesc[500]{Networks~Network performance analysis}

\keywords{p2p, libp2p, NAT traversal, hole punching, STUN, TURN, ICE}


\maketitle

\section{Introduction}
\label{sec:introduction}

Services like Facebook, TikTok, YouTube, Netflix, and Amazon have been criticized due to the potential risks of serving as single points of technical or organizational failures, becoming data monopolies, or acting as gatekeepers~\cite{Trinh2022, bommelaer2019global}. The resulting compromised privacy and loss of self-sovereign control over one's data poses significant risks. The vast amounts of personal information collected by these services have put them in a position of unprecedented power to influence users' lives and societies. Unlike other institutions that are legitimized by democratic principles, these influential services lack the same legitimization~\cite{mcintosh2018we}.

A growing movement, colloquially referred to as the ``Decentralized Web''~\cite{vojir2022} is challenging the power hierarchy of the current Web by eliminating intermediaries in web transactions using peer-to-peer (P2P) architectures. P2P replaces the current client-server paradigm and introduces the challenge of opening connections unilaterally and across restrictive Network Address Translators (NATs).


%

NAT technology was initially developed to address IPv4 address space depletion and is widely used by Internet Service Providers (ISPs) and corporate networks to enable devices to share a single public IP address. This usually results in unobstructed access from within a local area network (LAN) to the Internet, but hinders incoming connections.

NAT traversal techniques, such as NAT ``hole punching'', have been developed to overcome the challenge of ubiquitous P2P connectivity~\cite{Ford2005PeertoPeerCA}. In its simplest form, NAT hole punching allows two peers behind NATs to establish a direct connection by simultaneously opening a connection to the public IP/Port combinations of each other. Traditional hole punching techniques like the ones used by WebRTC~\cite{rfc8825}, rely on a centralized signaling server to facilitate the synchronization. This provides an opportunity to decentralize infrastructure and bolster a network's resilience against targeted attacks and censorship efforts. Decentralization alleviates the burden of operating and maintaining such infrastructure. As a result, the Direct Connection Upgrade through Relay protocol (DCUtR)~\cite{Seemann2022-jl} was designed. DCUtR draws inspiration from related protocols, including STUN, TURN, and ICE~\cite{rfc8489, rfc8656, rfc8445} but achieved independence from centralized infrastructure. 

In this paper, we specifically focus on the decentralized InterPlanetary File System (IPFS)~\cite{Trautwein2022,benet2014ipfs} network with its around 7k online peers\footnote{\url{https://probelab.io/ipfs/kpi/}}. 
IPFS is built upon the libp2p peer-to-peer networking stack, which includes the DCUtR protocol. At any given time, there are typically around 100k~\cite{maxinden} compatible libp2p nodes online beyond IPFS.
We evaluate DCUtR's real-world performance using five central hypotheses (see Section~\ref{sec:dcutr-hypotheses}). First, we establish its baseline efficacy (H1) and verify that its behavior is independent of relay characteristics (H2). Next, we examine both the efficiency and the transport-agnostic nature of its synchronization mechanism (H3a, H3b). Finally, we assess the practical impact of its built-in optimizations (H4). In a large-scale measurement campaign, we provide empirical evidence to validate or challenge these core hypotheses.

Our contributions are as follows:
(1) We present the first large-scale measurement study on a decentralized NAT traversal protocol in a production P2P network, 
(2) we empirically challenge long-the standing belief that UDP-based protocols are inherently superior for NAT traversal, 
(3) we validate DCUtR's core design; and
(4) we provide a unique, large-scale and open access dataset for the research community. 

\section{The Challenge of NAT for Peer-to-Peer Connectivity}
\label{sec:network-address-translation}


Network Address Translation (NAT) is a prevalent technique used in network gateways to modify IP address information; primarily, to mitigate IPv4 address exhaustion~\cite{rfc1631}. The most common variant, Network Address Port Translation (NAPT), also known as Port Address Translation (PAT), allows multiple devices within a private network to share a single public IP address~\cite{rfc2663}. The NAPT device maintains a dynamic translation table, mapping an internal source `(IP:port)` tuple to a public-facing `(IP:port)` tuple for each outbound connection.

This mapping and filtering mechanism allows outgoing connections but blocks unsolicited inbound traffic, which obstructs peer-to-peer (P2P) networking. A gateway only forwards an incoming packet if it matches a recent outgoing connection; otherwise, the packet is dropped. Hence, this mechanism requires specialized NAT traversal techniques to establish direct P2P links.


\subsection{Taxonomy of NAT Behavior}

The viability of any NAT traversal technique depends fundamentally on the NAT's mapping and filtering behavior ~\cite{rfc5382}, as standardized in RFC 4787~\cite{rfc4787}. Rather than examining these properties in isolation, it is more useful to consider how they combine to form distinct NAT archetypes, each with direct implications for P2P connectivity.

The most P2P-friendly NATs are known as \textbf{Cone NATs}. Their defining feature is \textit{Endpoint-Independent Mapping (EIM)}, where the NAT assigns a stable public endpoint `(IP:port)` to an internal endpoint, reusing this mapping for all subsequent outbound connections, regardless of the destination. This predictability is essential for hole punching, as a peer can learn its public endpoint once and reliably share it with others. Cone NATs differ primarily in their filtering policies. \textit{Full Cone NATs} (Endpoint-Independent Filtering) are the most permissive, accepting inbound traffic from any external source once a mapping exists. \textit{Restricted Cone NATs} (Address-Dependent or Address and Port-Dependent Filtering) enforce stricter rules, allowing inbound packets only from external endpoint `(IP:port)` to which the internal peer has recently transmitted traffic.

At the opposite end of the spectrum are \textbf{Symmetric NATs}, which pose the greatest challenge for P2P connectivity. A symmetric NAT uses \textit{Address and Port-Dependent Mapping (APDM)}, generating a unique public mapping for each distinct destination endpoint. As a result, the public address that a peer observes when contacting a discovery server differs from the one created when attempting to contact another peer. This unpredictability makes traditional hole punching ineffective, as the advertised public endpoint is no longer valid for new connections. This behavior, combined with restrictive filtering, often leaves relay-based solutions the only practical means of establishing peer-to-peer connectivity.

\subsection{NAT Traversal Techniques}

The de facto standard for handling the complexities of NAT is the Interactive Connectivity Establishment (ICE) framework~\cite{rfc5245, rfc8825}, which coordinates several techniques to find the most efficient communication path between peers. ICE systematically attempts connectivity using the following steps:

\textbf{Address Discovery with STUN:} Peers first query a centralized \textit{Session Traversal Utilities for NAT (STUN)} server~\cite{rfc8489}. The STUN server reflects the peer's public IP address and port as observed from the public internet (its "server reflexive" address) and helps classify the behavior of the NAT.

\textbf{Direct Connection via Hole Punching:} Equipped with their respective public addresses, peers exchange this information through a signaling channel (typically, a centralized service) and attempt to establish a direct connection using \textit{hole punching}~\cite{rfc5128}. This technique involves sending simultaneous connection-request packets to each other, taking advantage of the temporary NAT mappings created by their own outbound traffic.

\textbf{Relay Fallback with TURN:} If direct connection attempts fail, most commonly because one or both peers are behind a Symmetric NAT, ICE falls back to relaying traffic through a \textit{Traversal Using Relays around NAT (TURN)} server~\cite{rfc8656}. Although TURN guarantees connectivity, it increases latency and requires the TURN server to handle the full bandwidth of the session.

\noindent The staged NAT traversal approach of ICE aims to overcome the barriers that NATs impose on peer-to-peer communication, by offering a range of solutions from ``best-effort'' (hole punching) to guaranteed connectivity (TURN relaying).

Other mechanisms, such as UPnP~\cite{rfc6970} for automatic port forwarding or Application-Level Gateways (ALGs)~\cite{rfc2663}, exist but are not universally reliable or often disabled due to security concerns\footnote{https://www.cisa.gov/news-events/news/home-network-security}. Therefore, the robustness of the modern P2P ecosystem, particularly in applications like WebRTC, relies on the availability of centralized STUN, TURN, and signaling infrastructure. This dependency introduces operational costs and single points of failure, which oppose the principles of fully decentralized systems. This limitation motivated the development and analysis of protocols such as DCUtR.

\section{The libp2p DCUtR Protocol}
\label{sec:dcutr}

The Direct Connection Upgrade Through Relay (DCUtR) protocol is based on \texttt{libp2p}\footnote{\url{https://libp2p.io}}, a modular peer-to-peer networking stack in which DCUtR is implemented. \texttt{libp2p} is used by major peer-to-peer networks, such as Ethereum~\cite{Buterin2013} and IPFS~\cite{Trautwein2022,benet2014ipfs}, since it provides foundational tools for peer discovery, connection establishment, stream multiplexing, and secure communication. Within libp2p, several precursor protocols play a crucial role in gathering the necessary information and establishing initial connectivity for hole punching. These are notably the \textbf{Identify}\footnote{\url{https://docs.libp2p.io/concepts/fundamentals/protocols/\#identify}}, \textbf{AutoNAT}\footnote{\url{https://docs.libp2p.io/concepts/nat/autonat/}}, and the \textbf{Circuit v2}\footnote{\url{https://docs.libp2p.io/concepts/nat/circuit-relay/}} relay protocol. To  understand our following measurement methodology and evaluation, we give a brief overview of these precursor protocols as well as the DCUtR protocol operations. We point the reader to~\cite{Seemann2022-jl} for a more comprehensive description of the DCUtR protocol.

\subsection{libp2p}
\label{sec:dcutr-libp2p}

Identify, AutoNAT, and Circuit v2 protocols allow peers to discover their external network addresses and reachability. Further, the protocols establish preliminary connections without sole reliance on centralized STUN and TURN servers.

\subsubsection{Identify}

The Identify protocol provides functionality similar to STUN, is lightweight, and all libp2p peers support it by default. Upon establishing a connection, whether direct or relayed, peers exchange messages that allow each side to learn its own externally observed public IP address(es) and port as seen by the remote peer. Identify leverages existing connections, thus avoiding the need for separate centralized STUN infrastructure, and can potentially offer more reliable address information, since it uses the same transport protocols that will be used for the subsequent direct connection.

\subsubsection{AutoNAT}
\label{sec:dcutr-autonat}
Complementing this, the AutoNAT protocol determines a peer's actual reachability at its observed addresses. A peer requests that other peers attempt to dial back to the address it advertises. A successful dial-back classifies the peer as public, whereas failure indicates a private peer that likely cannot receive unsolicited inbound connections. On top of AutoNAT, a libp2p host also attempts to establish port mappings using Universal Plug and Play (UPnP) and/or the Port-Mapping Protocol (PMP). However, this mechanism is only advisory, since: \one routers may not reliably report the mapping's status, and \two mappings may silently expire at any time. AutoNAT provides the authoritative signal by informing the libp2p host which of its addresses are actually dialable from the outside.

\subsubsection{Circuit v2}
\label{sec:dcutr-circuit-v2}
For private peers, the Circuit v2 protocol provides lightweight relaying services, primarily for signaling and coordination tasks such as those required by DCUtR. Crucially, Circuit v2 is designed to minimize resource usage, imposing negligible processing and bandwidth costs on relays; unlike traditional TURN relays, which forwards all application traffic. Circuit v2 achieves this efficiency by requiring private peers to obtain reservations from relays, with strict limits on the number, duration, and data volume of relayed connections. This design enables most public libp2p peers to serve as relays without incurring significant resource burden. The establishment of this initial relayed connection via Circuit v2 is the precursor from which DCUtR attempts to upgrade to a direct connection.

\subsection{The DCUtR Protocol}
\label{sec:dcutr-protocol}

\begin{figure}
    \centering
    \includegraphics[width=\linewidth]{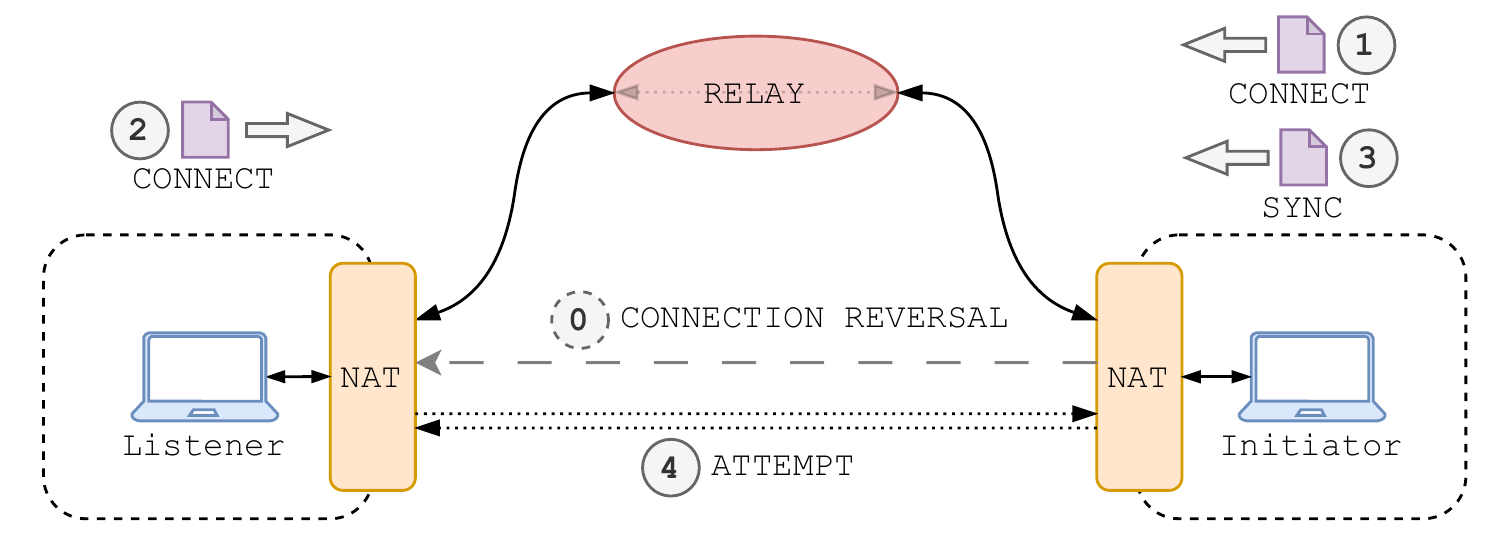}
    \caption{DCUtR protocol flow diagram}
    \label{fig:dcutr-flow-chart}
\end{figure}

The DCUtR protocol attempts to upgrade an existing relayed libp2p connection to a direct one. The process involves an \textbf{initiator}, the peer that accepted the relayed connection, and a \textbf{listener}, which waits for the initiator to begin the DCUtR exchange. Figure~\ref{fig:dcutr-flow-chart} depicts the protocol exchange which proceeds as follows. The setup begins with the initiator holding a valid reservation at any relay in the network. The listener can then use that relay to establish a ``limited'' relayed connection to start the exchange of Identify and DCUtR messages. Before attempting hole punching, the initiator may attempt a ``Connection Reversal'': if the listener appears public based on the Identify data obtained over the relayed connection, the initiator directly dials the listener's advertised addresses~\circled{0}. If this succeeds, a direct connection is formed without performing any hole punching. If Connection Reversal fails, or if the listener is private, DCUtR proceeds with the full hole punching workflow:

\textbf{Address Exchange:} The initiator sends a \texttt{CONNECT} message~\circled{1} containing its candidate public addresses (from Identify) to the listener via the relay. The listener responds with its own \texttt{CONNECT} message~\circled{2} containing its addresses. The initiator also uses this exchange to measure the relayed path round-trip time (RTT).

\textbf{Synchronization:} The initiator then sends a SYNC message~\circled{3} to the listener over the relay and waits half the measured RTT before initiating dialing.

\textbf{Dialing:} Upon receiving \texttt{SYNC}, the listener immediately attempts to establish a direct connection to the initiator's advertised addresses. After waiting half an RTT, the initiator also dials the listener's addresses~\circled{4}. This timing aims for near-simultaneous packet arrival at both NATs, increasing the likelihood of creating compatible NAT mappings.

\noindent Transport-specific mechanisms, such as TCP Simultaneous Open or QUIC's approach of using dummy UDP packets to establish NAT state, are leveraged. If the first synchronized attempt fails, the synchronization and synchronized dialing steps are retried twice. Once a direct connection is successfully established, the relayed connection is closed. The DCUtR coordination phase is intentionally lightweight, typically requiring only two network round trips and exchanging fewer than 500 bytes per direction over the relay. A detailed sequence diagram appears in Appendix~\ref{appendix:dcutr-sequence-diagram}. A comprehensive description is provided by Seemann et al.~\cite{Seemann2022-jl}.
DCUtR is shipped as a default component in the main libp2p implementations (go-libp2p, rust-libp2p, and js-libp2p) and is therefore active in all applications built on these stacks, including IPFS and Ethereum nodes.

\subsection{ICE/STUN/TURN Architectural Comparison} 
\label{sec:dcutr-vs-ice}

DCUtR departs from the ICE framework (Section~\ref{sec:network-address-translation}) which depends on dedicated STUN and TURN servers. DCUtR on the other hand leverages the libp2p's Identify protocol and ephemeral Circuit~v2 relays. Unlike the high-bandwidth fallback provided by TURN, DCUtR's relays are limited to signaling, meaning the system trades guaranteed connectivity for reduced infrastructure overhead. DCUtR also streamlines the handshake process using RTT-based synchronized dialing rather than systematic candidate checks and extends native support to both TCP and QUIC. This architectural pivot toward permissionless P2P networking informs the empirical analysis that follows.

\subsection{Protocol Hypotheses}
\label{sec:dcutr-hypotheses}

We formulate a set of testable hypotheses to examine DCUtR's real-world performance and behavior.
Each hypothesis reflects either a core design goal of the protocol or a known challenge in NAT traversal. The measurement campaign described in Section~\ref{sec:measurement} provides the empirical data needed to validate or refute these claims.


\pb{Viability and Efficacy.} The most fundamental requirement for any new protocol is basic efficacy. The public internet is a challenging environment for peer-to-peer communication, with a significant fraction of NAT devices (e.g., Symmetric NATs) intentionally designed to be exceptionally difficult to traverse~\cite{acunto2009}. Hole punching is a best effort technique with no guaranteed success. To consider DCUtR a viable connection strategy, we set a conservative but essential viability benchmark above 50\% and require that it must succeed more often than it fails.

    \textbf{H1:} We hypothesize that DCUtR achieves a success rate greater than 50\% for the hole punching stage, demonstrating its fundamental viability as a robust and effective decentralized NAT traversal solution in a heterogeneous real-world environment. Evaluated in Section~\ref{sec:analysis-efficacy}.

\pb{Relay Independence.} A defining property of the protocol is its decentralized architecture, which allows any public peer to serve as a signaling relay. This design is robust only if success does not depend on a small subset of privileged, high-performance, or strategically located relays.

    \textbf{H2:} We hypothesize that the success of a DCUtR hole punch is largely independent of the network characteristics of the relay used. This would validate the architectural choice of allowing any public peer to act as a relay in a permissionless network. Evaluated in Section~\ref{sec:analysis-relay-independence}.

\pb{Effective Synchronization.} DCUtR relies on an RTT-based synchronization mechanism to coordinate the hole punch. If effective, this synchronization mechanism should yield two outcomes: \one high efficiency: most successes occurring on the first attempt; and \two transport-agnostic behavior: mitigating the timing challenges that historically made TCP hole punching significantly harder than UDP-based methods. These considerations lead to two related hypotheses, evaluated in Section~\ref{sec:effective-synchronization}:

    \textbf{H3a:} We hypothesize that the DCUtR synchronization mechanism is highly efficient, with successful traversals predominantly occurring in the initial attempt.
    
    \textbf{H3b:} We hypothesize that DCUtR's performance is not fundamentally tied to the underlying transport protocol, achieving comparable success rates for both TCP and QUIC and challenging the conventional assumption of UDP's inherent superiority.

\pb{Optimization Effectiveness.} The protocol  includes ``Connection Reversal'' as a fast-path optimization to bypass the full hole-punching procedure when a peer already has a valid port mapping (e.g., via UPnP). An empirical evaluation must confirm whether this optimizations provides meaningful benefit in practice.

    \textbf{H4:} We hypothesize that the Connection Reversal mechanism is an effective optimization that can substantially increase the likelihood of establishing direct connections for peers with favorable NAT configurations and avoiding the overhead of the full hole-punching exchange. Evaluated in Section~\ref{sec:optimization-effectiveness}.

\section{Measurement Methodology \& Campaign}
\label{sec:measurement}

The objective of our measurements is to assess the performance of the DCUtR protocol, with a specific focus on validating or refuting the hypotheses in Section~\ref{sec:dcutr-hypotheses} and to gather insights into areas for protocol improvement. To achieve these objectives, we conducted a measurement campaign designed to comprehensively assess the performance of the DCUtR protocol. This section covers the measurement architecture, details on our measurement campaign, and the dataset of our research.

\subsection{Methodology}
\label{sec:methodology}

\begin{figure}[]
    \centering
    \includegraphics[width=\linewidth]{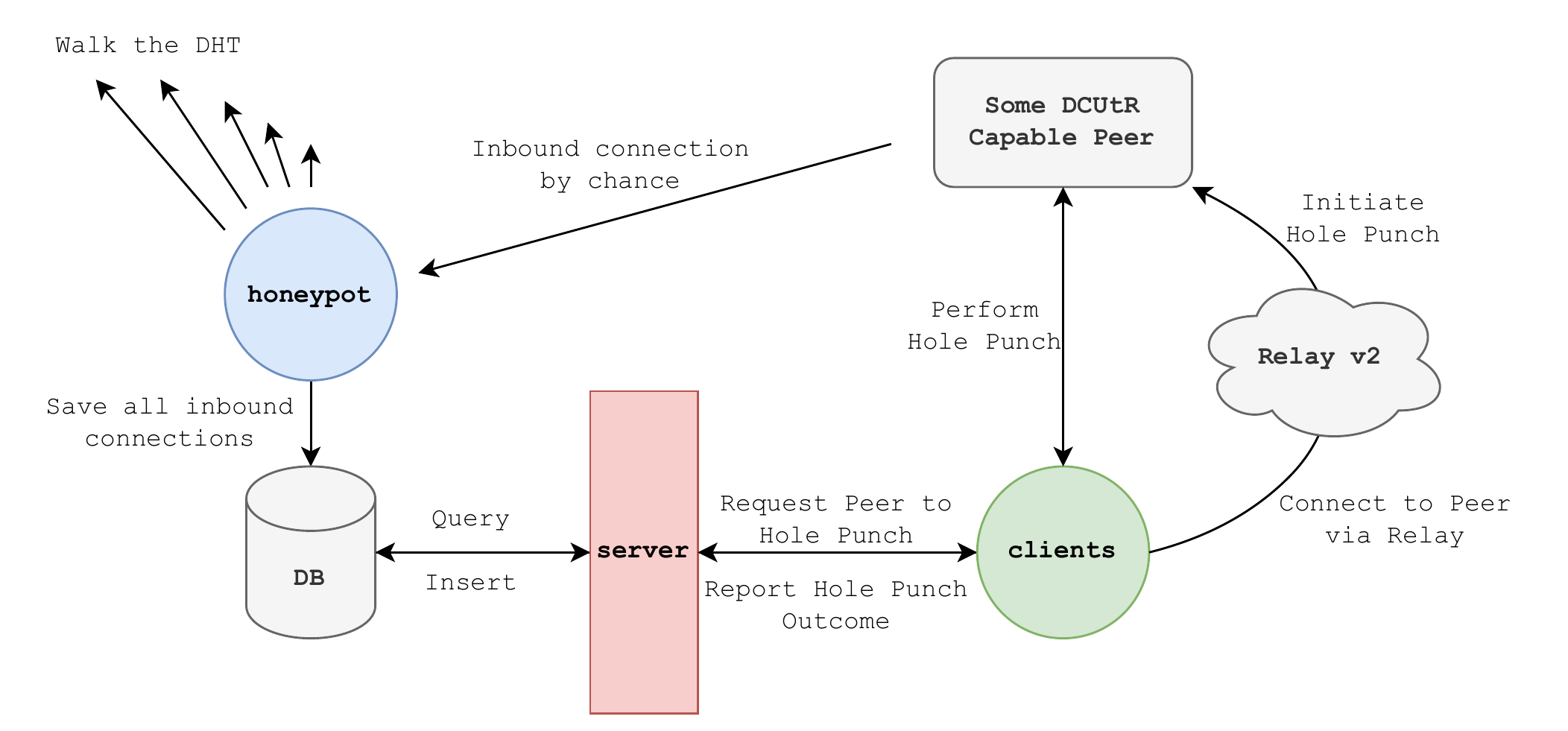}
    \caption{Measurement infrastructure architecture. The central components \textbf{honeypot}, \textbf{server}, and \textbf{clients} allow us to detect, serve and hole punch DCUtR-capable remote peers.}
    \label{fig:architecture}
\end{figure}

A central challenge in evaluating the performance of the DCUtR protocol is discovering peers that actually support DCUtR. For a meaningful assessment of NAT traversal capabilities, target peers must be located behind NATs, which makes them unreachable from the public internet. This characteristic complicates their discovery in decentralized, permissionless peer-to-peer networks as they typically, and specifically in the case of the IPFS network, lack a central registry of participants or of the protocols they support.

Standard peer discovery mechanisms, such as querying public Distributed Hash Tables (DHTs)~\cite{Trautwein2022}, are insufficient for this task. By default, only publicly reachable peers announce themselves to the DHT. NATed peers, precisely those required for our measurement campaign, do not. To overcome this limitation, we introduce a \textbf{honeypot}.

\subsubsection{Honeypot}

The honeypot is a DHT server peer designed to be highly stable and to announce itself to the network by slowly crawling it~\cite{Trautwein2022}. This behavior increases the likelihood that other peers insert the honeypot into their routing tables. As a result, server peers increasingly redirect client traffic to the honeypot, thereby raising the number of inbound connections it receives. The honeypot tracks all inbound connections from peers that satisfy both of the following conditions: \one the peer supports the DCUtR protocol \two the peer is reachable only through relay addresses, indicating it is behind a NAT. ``Tracking all inbound connections'' means that these connections are stored in the shared database, which is subsequently queried and served by the \textbf{server} component.

\subsubsection{Server}

The server component exposes a Google Remote Procedure Call (gRPC) API that allows clients to query for NATed and DCUtR-capable peers previously discovered by the honeypot. It is also responsible for tracking the results of hole-punching attempts and provides a centralized interface through which clients retrieve information about potential peers to hole punch. A key element of our methodology is the ``protocol filter'', a mechanism that enables us to isolate and evaluate the performance of specific transport protocols (TCP and QUIC) independently. When a client queries the server for a peer to hole punch, the server randomly assigns a ``protocol filter'', which the client should apply to the hole punch. For example, under a  ``TCP'' protocol filter, the client will \one announce only its public TCP listening address to the remote peer over the relay and \two attempt the hole punch using only the remote peer's public TCP Multiaddress. This ensures that both endpoints rely exclusively on the designated transport protocol, despite potentially listening on additional ones.

\subsubsection{Clients}

Clients periodically query the server for peers to probe. Using the returned information, they perform hole-punching attempts and report their results back to the server. Clients connect to each peer via its advertised relay address, thereby acting as the previously introduced \textit{listener}. They then wait for the remote peer, \textit{initiator}, to initiate the hole punch protocol. This component generates the empirical data for the measurement campaign, enabling the server to collect and aggregate results. The client is written in Go and is available in two forms: a command-line interface (CLI) and a graphical user interface (GUI). The GUI version was developed to simplify onboarding of participants in the measurement study. The source code is openly available at \href{https://github.com/libp2p/punchr}{https://github.com/libp2p/punchr}.

\subsubsection{Interplay}

Figure~\ref{fig:architecture} illustrates the interaction between the components. The honeypot continuously crawls or ``walks'' the DHT to increase the likelihood that NATed, DCUtR-capable peers connect to it. These inbound connections are stored in a database, which is then queried and served to clients. Clients use the provided relay address(es) to connect to the remote peer and initiate the DCUtR hole-punching process. After either \one successfully establishing direct connection, \two three consecutive DCUtR failures, or \three a timeout in establishing the connection, the client reports the outcome to the server. The server then persists these results for subsequent analysis.


\subsection{Study}
\label{sec:campaign}
\begin{figure*}[]
    \centering
    \begin{subfigure}[t]{0.49\textwidth}
        \refstepcounter{subfigure}
        \begin{minipage}[c]{0.01\linewidth}
            (\thesubfigure) 
        \label{fig:geo-clients}
        \end{minipage}%
        \begin{minipage}[c]{1\linewidth}
            \centering
        \includegraphics[width=0.9\linewidth]{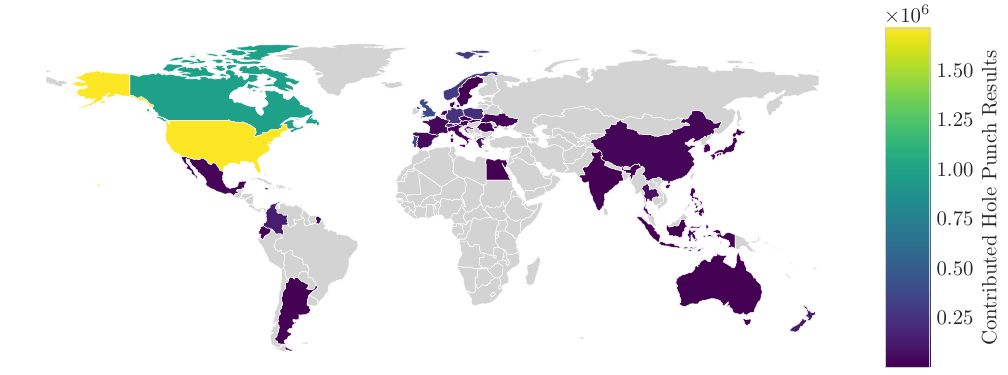}
        \end{minipage}
            
    \end{subfigure}
    \hfill
    \begin{subfigure}[t]{0.49\textwidth}
        \refstepcounter{subfigure}
        \begin{minipage}[c]{0.01\linewidth}
            (\thesubfigure) 
            \label{fig:geo-remote-peers}
        \end{minipage}%
        \begin{minipage}[c]{1\linewidth}
            \centering
        \includegraphics[width=0.9\linewidth]{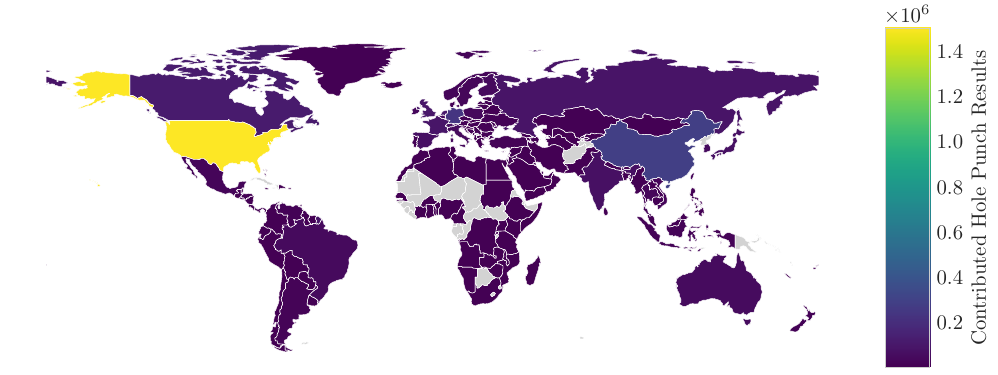}
        \end{minipage}
        \end{subfigure}
    
    \caption{(a) Geographic distribution of controlled client peers in the measurement study that contributed hole punch results. (b) Geographic distribution of remote peers that interacted with the IPFS network that contributed hole punch results in the measurement campaign.}
\end{figure*}

Our measurement study ran from December 1, 2022, to January 1, 2023. We announced the study both publicly in libp2p and IPFS community channels. Interested participants were invited to sign up via a form that collected general information about the network in which they primarily intended to run the client. The participants received an API key from us, which allowed us to link the collected data points with their questionnaire responses. Importantly, we imposed no restrictions on client mobility: participants were free to move their client between networks at their discretion. This introduces a challenge in the data analysis, as data points for a given client must be associated with the specific network in which they occurred. We discuss this challenge in Section~\ref{sec:network-identification}, but we believe that allowing such flexibility was crucial to maximizing participation.

Participation did not require registration. Users could download the source code, run the clients themselves, and report results directly to our server. In these cases, the server generated a random API-Key for the participant. Any detected abuse by non-registered clients could be excluded from the dataset; however, we did not observe any such behavior.

\subsection{Dataset}
\label{sec:dataset}

\begin{table*}
    \centering
    \caption{The possible outcomes of an individual hole punch result as reported by the clients. Each hole punch results can consists of up to three ``attempt'' data points. Their outcomes are listed in table \ref{tab:hpa-outcomes}}
    \label{tab:hpr-outcomes}
    \vspace{-0.3cm}
    \begin{tabularx}{\textwidth}{lX}
    \hline
    \hline
         Outcome & Description \\ \hline
         \textbf{\texttt{UNKNOWN}} & There is no information why and how the hole punch completed. \\
         \textbf{\texttt{NO\_CONNECTION}} & The client could not connect to the remote peer via any of the provided multi addresses. \\
         \textbf{\texttt{NO\_STREAM}} & The client could connect to the remote peer via any of the provided multi addresses but no \texttt{/libp2p/dcutr} stream was opened within 15s. That stream is necessary to perform the hole punch. \\
         \textbf{\texttt{CONNECTION\_REVERSED}} & The client only used one or more relay multi addresses to connect to the remote peer, the \texttt{/libp2p/dcutr} stream was not opened within 15s, and we still end up with a direct connection. This means the remote peer successfully reversed it. \\
         \textbf{\texttt{CANCELLED}} & The user stopped the client (also returned by the rust client for quic multi addresses). \\
         \textbf{\texttt{FAILED}} & The hole punch was attempted multiple times but none succeeded OR the \texttt{/libp2p/dcutr} was opened but we have not received the internal start event OR there was a general protocol error. \\
         \textbf{\texttt{SUCCESS}} & Any of the hole punch attempts succeeded. \\
    \hline
    \hline
    \end{tabularx}
\end{table*}

\begin{table*}
    \centering
    \caption{The possible outcomes of an individual hole punch attempt as reported by clients. Each hole punch results can consists of up to three ``attempts'' data points with the possible outcomes listed below. The outcome of the attempts informs the overall outcome of the result in table~\ref{tab:hpr-outcomes}}
    \label{tab:hpa-outcomes}
    \vspace{-0.3cm}
    \begin{tabularx}{\textwidth}{lX}
    \hline
    \hline
         Outcome & Description \\ \hline
         \textbf{\texttt{UNKNOWN}} & There was no information why and how the hole punch attempt completed. \\
         \textbf{\texttt{DIRECT\_DIAL}} & The connection reversal from our side succeeded (should never happen). \\
         \textbf{\texttt{PROTOCOL\_ERROR}} & This can happen if e.g., the stream was reset mid-flight. \\
         \textbf{\texttt{CANCELLED}} & The user stopped the client. \\
         \textbf{\texttt{TIMEOUT}} & We waited for the internal start event for 15s but timed out. \\
         \textbf{\texttt{FAILED}} & We exchanged \texttt{CONNECT} and \texttt{SYNC} messages on the \texttt{/libp2p/dcutr} stream but the final direct connection attempt failed; the hole punch was unsuccessful. \\
         \textbf{\texttt{SUCCESS}} & We were able to directly connect to the remote peer.  \\
    \hline
    \hline
    \end{tabularx}
\end{table*}

In total, we tracked over 6.25 million hole punches across 212 API keys while registering 148 sign-ups. Figures \ref{fig:geo-clients} and \ref{fig:geo-remote-peers} show that the clients were deployed in 39 countries and hole-punched remote peers in 167 countries. The maps also show the number of data points contributed from each country. While the sample distibution is skewed toward the U.S., our dataset captures globally distributed measurements.

Each time a client completes a hole punch probe, it reports a \textbf{hole punch result}, which may contain multiple \textbf{hole punch attempts}, as each hole punch is tried up to three times. Each attempt has an individual ``outcome'' in addition to the overall result. Tables~\ref{tab:hpr-outcomes} and~\ref{tab:hpa-outcomes} list the possible outcomes for a hole punch \textit{result} and \textit{attempt}, respectively. Understanding these outcomes is important for interpreting our evaluation. 

Each data point includes the following information: the client and remote peer identifiers, all IP addresses and ports that the client is listening on, the set of addresses used to connect to the remote peer (typically a single relay address), the set of addresses used to directly connect to the peer, any open connections to the remote peer after the hole punch, and any active port mappings on the client side. We make our dataset available under the CC BY-SA license with the following IPFS content identifier:

\begin{center}
    \footnotesize
    \href{https://bafybeia7sq3nfd7c4obcy7ahjvnoka7ujdiob33r7rqyeycgicdt3iknki.ipfs.dweb.link/}{\texttt{bafybeia7sq3nfd7c4obcy7ahjvnoka7ujdiob33r7rqyeycgicdt3iknki}}
\end{center}

\noindent A detailed description of the dataset structure and content is found in Appendix~\ref{appendix:dataset}. Personally identifiable information has been anonymized.

\subsubsection{Limitations}
\label{sec:dataset:limitations}

We acknowledge several methodological limitations that may affect interpretation of the evaluation results.

\pb{Client Sampling.} Our client fleet was recruited from volunteers in the IPFS and libp2p communities. This group is likely more technically proficient and may operate on more stable or less restrictive networks than the general P2P user base. This selection bias may lead to an overestimation of the protocol's success rate (H1) and may limit the generalizability of our findings.

\pb{Discovery Bias.} Our measurement setup relies on a 'honeypot' that discovers remote, NATed peers by accepting inbound relayed connections. This approach systematically excludes peers behind highly restrictive NATs or firewalls that prevent even relayed connections. Consequently, the set of remote peers in our study is pre-filtered for a baseline level of reachability, which may inflate the observed hole-punching success rate.

\pb{NAT Type Classification.} Our measurement architecture observes only the
\textit{outcome} of hole-punch attempts (success or failure), not the
internal mapping and filtering behavior of remote NAT devices.
Classifying NAT types as defined in RFC~4787~\cite{rfc4787} requires
dedicated multi-server probing techniques~\cite{rfc5780} that were
outside the scope of our campaign. This also extends to Carrier-Grade NAT (CGNAT) deployments, which introduce an additional NAT layer and frequently employ EDM translation~\cite{Richter2016}; our data cannot distinguish single-NAT from CGNAT peers. As a result, we cannot and do not attribute individual failures to specific NAT configurations.


\section{Evaluation}
\label{sec:evaluation}


In this section, we evaluate the data collected during our study to empirically test the hypotheses outlined in Section~\ref{sec:dcutr-hypotheses}. We describe our data-preparation steps, analyze the protocol's overall success rate and its dependence on network and transport factors, and assess its efficiency.




\subsection{Data Preparation -- Network Identification}
\label{sec:network-identification}

Our analysis of DCUtR's performance begins with identifying the individual networks from which clients operated. Because clients are mobile, it is necessary to segment each client's data by the network environment in which the measurements occurred. This avoids conflating results from different network conditions (e.g., distinct NAT devices and policies). We identify individual networks using two assumptions: \one data points belong to the same network if the client reports the same public IP address, and \two if a client's public IP changes but remains within the same Autonomous System (AS), we treat the measurements as coming from the same network provided that the set of locally assigned private IP addresses remains unchanged. 
We acknowledge that the second heuristic may conflate distinct networks when a client moves between locations served by the same AS, since consumer NAT devices within a single provider can assign similar private address ranges.
Applying this method, we identify $859$ distinct networks that the clients operate in (no two clients operated from the same network) and $86,769$ networks for remote peers.

\begin{figure}[]
    \centering
    \begin{subfigure}[t]{0.23\textwidth}
        \refstepcounter{subfigure}
        \begin{minipage}[c]{0.05\linewidth}
            (\thesubfigure) 
        \label{fig:networks-distribution}
        \end{minipage}%
        \begin{minipage}[c]{1\linewidth}
            \centering
            \includegraphics[width=0.9\linewidth]{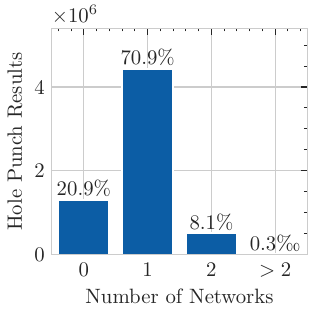}
        \end{minipage}
            
    \end{subfigure}
    \hfill
    \begin{subfigure}[t]{0.24\textwidth}
        \refstepcounter{subfigure}
        \begin{minipage}[c]{0.05\linewidth}
            (\thesubfigure) 
        \label{fig:client-networks}
        \end{minipage}%
        \begin{minipage}[c]{1\linewidth}
            \centering
            \includegraphics[width=0.9\linewidth]{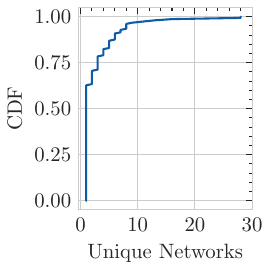}
        \end{minipage}
    \end{subfigure}
    \caption{(a) Network identifications per hole punch result. Most could be linked to a single network based on the reported IP addresses that the client listens on. (b) CDF of the number of unique networks per client, showing that over $60\%$ of clients operated from a single network throughout the study.}
\end{figure}

Figure~\ref{fig:networks-distribution} shows that our network identification process successfully associated approximately 4.43 million data points ($70.9\%$ of the 6.25 million reported hole-punch results) with a single, unambiguous client network. In contrast, $20.9\%$ of data points were associated with ``0 public networks,'' meaning the client did not advertise a public IP address. This can occur when no relay reservation was in place yet or when the Identify protocal is unable to determine a public address. An additional $8.1\%$ of data points correspond to clients that reported multiple public IPs across different networks, which can arise when a client is assigned more than one public IP address.
Figure~\ref{fig:client-networks} further illustrates that the majority (over $60\%$) of clients operated exclusively within a single network. At the other extreme, one highly mobile client was observed operating from 28 distinct networks during the measurement period.

For the following analysis, we only consider hole punch results that we could unambiguously associate with a single network.
The excluded data points represent fundamental failures in the broader connectivity stack upon which DCUtR depends. While our subsequent analysis focuses on the efficacy of the hole-punching stage itself, this initial failure rate underscores the multi-layered challenges inherent in establishing P2P connectivity under real-world conditions.
We revisit this and other limitations in Section~\ref{sec:discussion}.

\subsection{Analysis}
\label{sec:analysis}

In the following sections, we evaluate the hypotheses from Section~\ref{sec:dcutr-hypotheses}.





\subsubsection{H1: Viability and Efficacy.}
\label{sec:analysis-efficacy}

\begin{figure*}[]
    \centering
    \begin{subfigure}[t]{0.46\textwidth}
        \refstepcounter{subfigure}
        \begin{minipage}[c]{0.001\linewidth}
            (\thesubfigure) 
            \label{fig:campaign-outcomes}
        \end{minipage}%
        \begin{minipage}[c]{1\linewidth}
            \centering
            \includegraphics[width=0.9\linewidth]{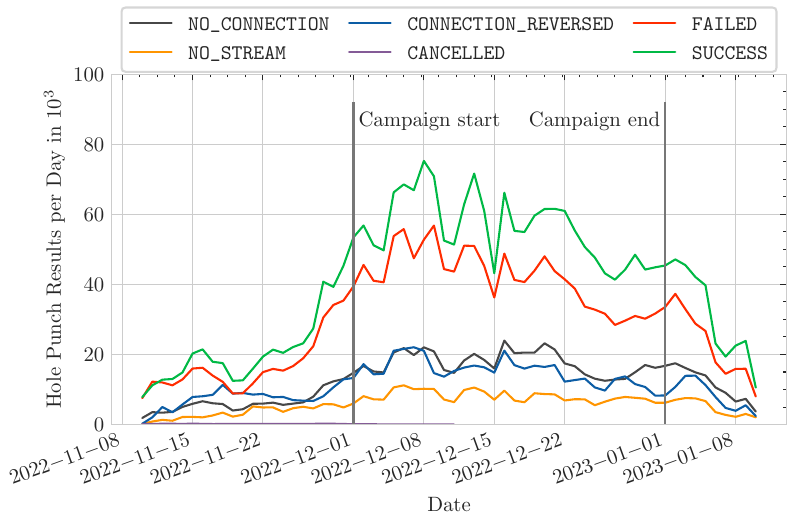}
        \end{minipage}
            
    \end{subfigure}
    \hfill
    \begin{subfigure}[t]{0.53\textwidth}
        \refstepcounter{subfigure}
        \begin{minipage}[c]{0.001\linewidth}
            (\thesubfigure) 
            \label{fig:network-success-rates}
        \end{minipage}%
        \begin{minipage}[c]{1\linewidth}
            \centering
            \includegraphics[width=0.9\linewidth]{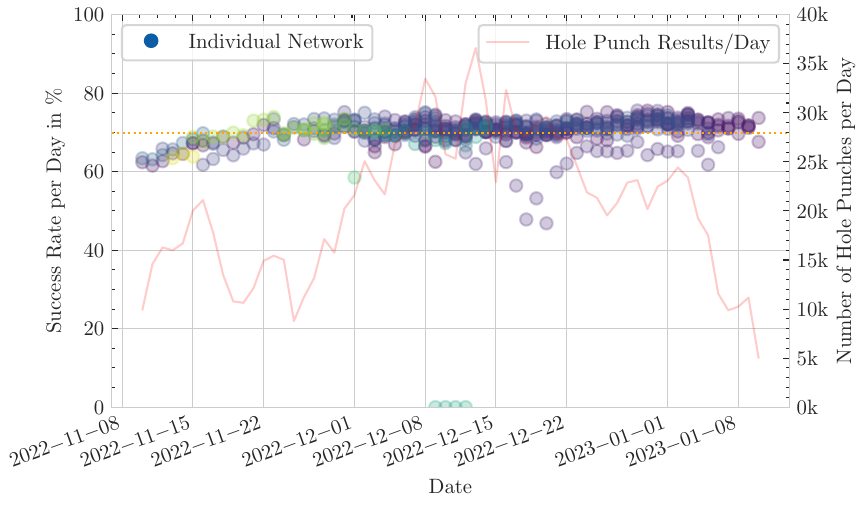}
        \end{minipage}
            
    \end{subfigure}
    \caption{(a) Reported hole punch results over the course of the duration of our measurement campaign split by their individual outcomes according to table~\ref{tab:hpr-outcomes}. (b) Daily success rates of hole punches for individual networks across the entire measurement period. The dashed orange line is the line of best fit across all success rates.}
\end{figure*}

To test our primary hypothesis regarding the viability of decentralized NAT traversal (H1), we first establish the baseline success rate of the DCUtR protocol across our large-scale deployment. Figure~\ref{fig:campaign-outcomes} shows the daily number of reported hole punch results grouped by outcome categories defined in Table~\ref{tab:hpr-outcomes}. The graph shows that the number of contributed hole punches increased after December 1, declined over Christmas, and tapered off following the distribution of the campaign termination notice in early January. This pattern aligns with the timing of our study and expected client availability during the holiday period (see Figures~\ref{fig:geo-clients} and~\ref{fig:geo-remote-peers}). 

To derive the success rate that a peer would experience when attempting to connect to a random peer in the IPFS network, we consider only hole-punch results that meet the following criteria: 
\one no port mapping was already in place, 
\two the client contributed more than 1,000 data points from the respective network, and 
\three the outcome is either \texttt{SUCCESS} or \texttt{FAILED} (see Table~\ref{tab:hpr-outcomes}), ensuring that a hole-punch attempt was actually performed.
The 1,000 data point threshold ensures statistical stability of per-network success rates by excluding transient or minimally contributing clients whose small sample sizes would introduce high variance into the aggregate metric.

Figure~\ref{fig:network-success-rates} shows the per-network success rates, visualized as shaded points averaged per day. The secondary y-axis indicates the number of data points that satisfy the filtering criteria, providing context for the sample size. 
A linear fit across all networks over the measurement period yields an average \textbf{conditional hole-punch success rate of $70\% \pm 7.1\%$}. This rate depends on successful relay reservation and public address discovery which themselves fail for approximately $29\%$ of attempts (see Section~\ref{sec:discussion}). Consequently, the end-to-end probability of establishing a direct connection from scratch is lower than the hole-punch success rate. This consistently high success rate across a diverse and global set of networks provides strong evidence in support of H1, demonstrating that DCUtR is an effective protocol for establishing direct P2P connections. We discuss the implications of this rate further in Section~\ref{sec:discussion}. However, we acknowledge that this number is not indicative of any individuals' experienced success rate. It instead gives a statistical average across the network and tells how likely it is for two random peers behind NAT devices to establish a direct connection using DCUtR.



Having established the viability of the DCUtR protocol with a consistent success rate (H1), we now quantify the tangible performance benefits for peers that successfully establish a direct connection. Figure~\ref{fig:latency-impact} illustrates the latency reduction achieved when upgrading from a relayed to a direct path. For each successful hole punch, we compute the ratio between the direct RTT to the remote peer and the RTT measured via the relay. For example, if the RTT through the relay is 1 s and the post-hole punch direct RTT is 0.7 s, the resulting ration is $70\%$. The figure shows that $50\%$ of peers experience a reduction to $70\%$ or less of their original RTT. Approximately $10\%$ of peers report a higher RTT after the hole punch (all points above $10^0$), while $90\%$ benefit from a reduced RTT. These results confirm that hole punching can significantly improve performance and enable delay-sensitive applications. Cases where the direct path is slower than the relayed path can occur if the relay is located on a high-speed internet backbone, while the direct peer-to-peer path traverses slower consumer-grade ISP networks.

\subsubsection{H2: Relay Independence}
\label{sec:analysis-relay-independence}

\begin{figure*}[]
    \centering
    \begin{subfigure}[t]{0.44\textwidth}
        \refstepcounter{subfigure}
        \begin{minipage}[c]{0.01\linewidth}
            (\thesubfigure) 
            \label{fig:latency-impact}
        \end{minipage}%
        \begin{minipage}[c]{1\linewidth}
            \centering
        \includegraphics[width=0.9\linewidth]{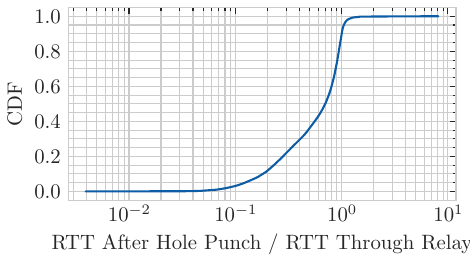}
        \end{minipage}
            
    \end{subfigure}
    \hfill
    \begin{subfigure}[t]{0.55\textwidth}
        \refstepcounter{subfigure}
        \begin{minipage}[c]{0.01\linewidth}
            (\thesubfigure) 
            \label{fig:rtt-dependence}
        \end{minipage}%
        \begin{minipage}[c]{1\linewidth}
            \centering
            \includegraphics[width=0.9\linewidth]{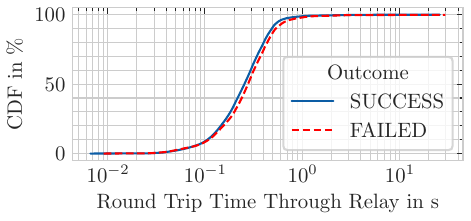}
        \end{minipage}
\end{subfigure}

    \caption{(a) The direct RTT as a fraction of the RTT through the relay. (b) RTT distributions for hole punches with the outcomes \texttt{SUCCESS} and \texttt{FAILED}. The sample sizes are $897,747$ and $684,376$, respectively.}
\end{figure*}

A key requirement for a decentralized relay system is independence from relay-specific characteristics (H2). We first test this hypothesis by investigating the influence of the round-trip time (RTT) through the relay on the hole-punch outcome. Figure~\ref{fig:rtt-dependence} compares two Cumulative Distribution Functions (CDFs) of RTTs to the remote peer via the relay: one for successful hole punches (solid blue) and one for failures (dashed red). While failed attempts exhibit slightly higher RTTs, suggesting a weak negative correlation, the overall difference is minor as the two CDFs remain closely aligned.



To further validate our hypothesis of relay independence (H2), we analyze whether the relay's location along the network path between the two peers affects success rates. 
The client-reported data include both the RTT to the relay and the RTT to the remote peer via the relay, enabling us to infer the relay's relative location.
If we define the RTT through the relay to the remote peer to be $100\%$ of the distance, then the RTT from the client to the relay reflects the fraction of the path the relay occupies. For example, if the RTT to the remote peer through the relay is $1\;\text{s}$ and the RTT to the relay is $700\;\text{ms}$, we define the relay as being $70\%$ away from the client to the remote peer. Figure~\ref{fig:relay-path-location-dependence} shows the success rate (blue) as a function of the relay's fractional path position to the remote peer (binned in $5\%$ increments), along with the number of hole punch results in which the relay was in that specific path location (red). The data shows that the success rates are largely independent of the location of the relay along the path. Most relay nodes fall near the midpoint between us and the remote peer, with a slight skew toward proximity to the client.

This skew is expected, since peers by default request reservations with two relays. Clients then dial remote peers via all available relay addresses. It is natural that the connection to the closer relay, and thus to the remote peer through it, succeeds first.




Taken together, these findings demonstrate only a minor effect from elevated RTTs and no measurable dependence on relay path location, providing compelling evidence for H2. The performance of DCUtR is thus robust to the position and latency of the facilitating relay, which is a critical characteristic for a permissionless, decentralized P2P system.

\subsubsection{H3: Effective Synchronization}
\label{sec:effective-synchronization}

\begin{figure*}[]
    \centering
    \begin{subfigure}[t]{0.49\textwidth}
        \refstepcounter{subfigure}
        \begin{minipage}[c]{0.01\linewidth}
            (\thesubfigure) 
            \label{fig:relay-path-location-dependence}
        \end{minipage}%
        \begin{minipage}[c]{1\linewidth}
            \centering
            \includegraphics[width=0.9\linewidth]{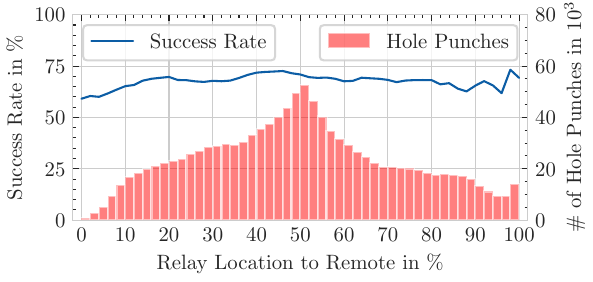}
        \end{minipage}
            
    \end{subfigure}
    \hfill
    \begin{subfigure}[t]{0.50\textwidth}
        \refstepcounter{subfigure}
        \begin{minipage}[c]{0.01\linewidth}
            (\thesubfigure) 
            \label{fig:rtt-accuracy}
        \end{minipage}%
        \begin{minipage}[c]{1\linewidth}
            \centering
            \includegraphics[width=0.9\linewidth]{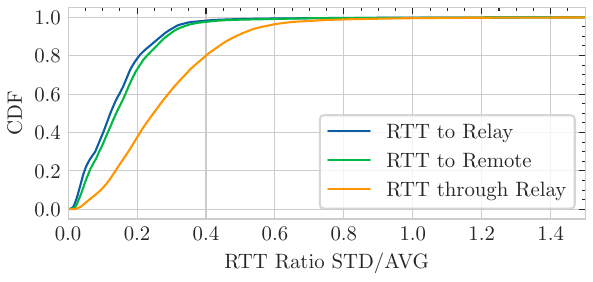}
        \end{minipage}
    \end{subfigure}
    \caption{(a) Success rate dependence on relay location along the path. (b) Ratio of the mean over the standard deviation of the RTT measurements}
\end{figure*}

The protocol's synchronized dialing mechanism is central to its design. We now test two related hypotheses: that synchronization is highly efficient (H3a) and sufficiently accurate to render the protocol transport-agnostic (H3b).
Effective synchronization requires that RTT measurements are accurate enough to coordinate simultaneous connection attempts. That is, packets from each peer must have departed their respective routers before packets from the other peer arrive. Figure~\ref{fig:rtt-accuracy} aggregates all RTT measurements collected during the measurement study. Each RTT measurement represents up to ten ping samples. We categorize the dataset into three RTT  scenarios: (1) client → relay, (2) client → remote peer via relay, and (3)  client → remote peer directly following a successful hole punch. The results show that for the critical relay-mediated RTTs (scenario 2), over $90\%$ of measurements exhibit a standard deviation less than half of the average RTT, indicating a substantial synchronization margin. As expected, latency measurements to the relay and to the remote peers exhibit similar distributions, with RTT variations generally smaller than those observed in the measurements involving the relay path to the remote peer.

\pb{Attempts (H3a).} With RTT measurement accuracy established, we assess H3a. Figure~\ref{fig:attempts-success} reveals that $97.6\%$ of successful hole punches complete on the very first attempt, with only $2.4\%$ requiring additional retries. This finding has direct implications for protocol optimization: if later attempts contribute only marginal gains, the DCUtR protocol may be simplified by reducing retries or adjusting the retry strategy. We discuss such protocol optimizations in Section~\ref{sec:optimizations}.





\pb{Transport Protocol Dependence (H3b).} We now evaluate H3b, which challenges the conventional assumption that UDP-based hole punching is inherently more effective than TCP-based techniques. This assumption is rooted in fundamental differences in how NATs handle TCP and UDP flows, and how NATs and firewalls process their traffic. We discuss these arguments in Section~\ref{sec:discussion}. To test transport agnosticism, we analyze the success rates for hole punches when explicitly restricted to either TCP or QUIC.



\begin{figure}[]
    \centering
    \begin{subfigure}[t]{0.24\textwidth}
        \refstepcounter{subfigure}
        \begin{minipage}[c]{0.05\linewidth}
            (\thesubfigure) 
        \label{fig:transport-success-share}
        \end{minipage}%
        \begin{minipage}[c]{1\linewidth}
            \centering
            \includegraphics[width=0.9\linewidth]{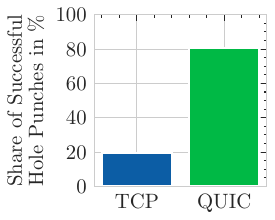}
        \end{minipage}
            
    \end{subfigure}
    \hfill
    \begin{subfigure}[t]{0.23\textwidth}
        \refstepcounter{subfigure}
        \begin{minipage}[c]{0.05\linewidth}
            (\thesubfigure) 
        \label{fig:transport-success-rate}
        \end{minipage}%
        \begin{minipage}[c]{1\linewidth}
            \centering
            \includegraphics[width=0.9\linewidth]{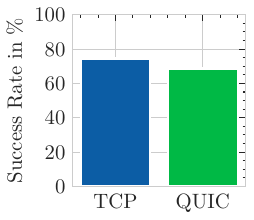}
        \end{minipage}
    \end{subfigure}

    \caption{(a) If the hole punch was successful, which transport was the final connection using (b) If the attempts were restricted to using a single transport, what are the respective success rates.}
\end{figure}

Figure~\ref{fig:transport-success-share} shows that when transport is unrestricted, roughly $\sim 80\%$ of resulting connections use QUIC. This reflects only that QUIC's connection establishment latency is lower than TCP's, a finding consistent with Liang et al.~\cite{liang2024}, and does not directly speak to hole-punch success. When we restrict hole-punch attempts to a specific transport protocol, Figure~\ref{fig:transport-success-rate} shows that both TCP and QUIC achieve success rates near $\sim70\%$, strongly validating H3b. DCUtR's deterministic synchronization mechanism effectively mitigates the traditional challenges of TCP traversal, demonstrating that the protocol is indeed transport-agnostic.

\subsubsection{H4: Effectiveness of the Connection Reversal Optimization}
\label{sec:optimization-effectiveness}

The DCUtR protocol includes Connection Reversal as a fast-path optimization. To test its real-world effectiveness (H4), we analyze the distribution of outcomes for peers that report an active port mapping. As noted in Section~\ref{sec:dcutr}, libp2p attempts to register port mappings with routers using UPnP or PMP. Hole punch result data points from our fleet of clients also comprise information about active port mappings in their local network. However, while these mappings are considered advisory due to potential router dishonesty or unannounced expirations, their presence is expected to influence hole punching outcomes. Specifically, a higher share of \texttt{CONNECTION\_REVERSAL} outcomes would be anticipated when an active port mapping is reported, as this allows a direct connection without the need for a full hole punch.

\begin{figure*}[]
    \centering
    \begin{subfigure}[t]{0.48\textwidth}
        \refstepcounter{subfigure}
        \begin{minipage}[c]{0.01\linewidth}
            (\thesubfigure) 
            \label{fig:attempts-success}
        \end{minipage}%
        \begin{minipage}[c]{1\linewidth}
            \centering
            \includegraphics[width=0.9\linewidth]{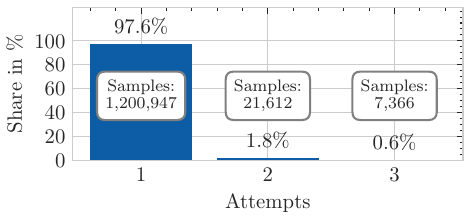}
        \end{minipage}
            
    \end{subfigure}
    \hfill
    \begin{subfigure}[t]{0.51\textwidth}
        \refstepcounter{subfigure}
        \begin{minipage}[c]{0.01\linewidth}
            (\thesubfigure) 
            \label{fig:port-mappings-outcomes}
        \end{minipage}%
        \begin{minipage}[c]{1\linewidth}
            \centering
            \includegraphics[width=0.9\linewidth]{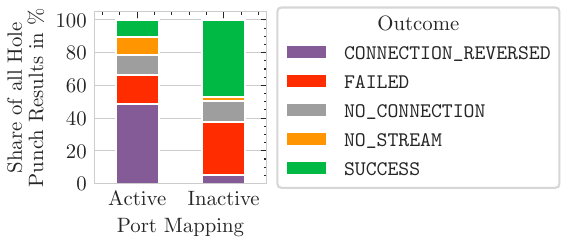}
        \end{minipage}
    \end{subfigure}

    \caption{(a) If successful, with which attempt succeeded the hole punch. (b) Influence of active port mappings on the hole punch outcome.}
\end{figure*}

Figure~\ref{fig:port-mappings-outcomes} shows the distribution of cases where a port mapping was active and inactive respectively. One can clearly see the significantly higher share of \texttt{CONNECTION\_REVERSAL} outcomes in the case of active port mappings on the clients' side of the connection. This observation empirically validates the importance of the connection reversal technique within the DCUtR protocol. If a direct connection can be established via a pre-existing port mapping, it is highly advantageous to prioritize this method over initiating a full hole punch. Connection reversal avoids the inherent complexities, timing sensitivities, and resource consumption associated with the hole punching procedure, making it a more efficient and reliable path to direct connectivity when applicable. This empirical validation of the Connection Reversal mechanism confirms H4 and underscores the value of including such preliminary checks to avoid the overhead of the full hole-punching procedure when possible.
We note that this analysis is based on port-mapping reports from our 212 volunteer clients, whose network configurations may not be representative of the broader population. 

\section{Discussion}
\label{sec:discussion}

Our empirical evaluation validates the core hypotheses regarding DCUtR's performance. The confirmation of relay independence (H2) is especially important for permissionless P2P systems, demonstrating that the protocol does not rely on a subset of privileged or well-positioned relays for correct operation. This robustness to the position and latency of the relay is essential when relay selection is random and relay quality is not guaranteed. Furthermore, our establishment of a contemporary, conditional baseline success rate of roughly $70\%$ across a large and diverse set of  networks over an extended period (H1) substantially expands on the scale and recency of prior work and provides a crucial benchmark for future optimizations and comparisons. Earlier studies, e.g. Halkes et al.~\cite{Halkes2011-og} in 2011 found that ``approximately $64\%$ of all peers are behind a NAT box or firewall which should allow hole punching to work, and more than $80\%$ of hole punching attempts between these peers succeed.'' To the best of our knowledge, only Guha et al.~\cite{Guha2005} reported ``in the wild'' hole-punching success rates. They reported an ``88\% average success rate for TCP connection establishment with NATs in the wild.'' However, their measurements comprised just 93 home NATs, which offers limited statistical power and lower representativeness than our considerably larger dataset. Moreover, all studies we identified rely on data that is considerably outdated. This may not accurately reflect the current internet landscape and underscores the need for renewed, large-scale measurement of NAT traversal in the current Internet.

\pb{Transport Protocol Dependence.} The validation of H3b (transport agnosticism) merits further discussion. TCP hole punching is widely considered difficult due to constraints of the three-way handshake and susceptibility to \texttt{RST} packets from stateful firewalls, which are well-documented~\cite{rfc5128,Guha2005}. Additionally, the standard Berkeley sockets API does not inherently support the simultaneous open and listening on the same port required for TCP hole punching. While \texttt{SO\_REUSEADDR} and \texttt{SO\_REUSEPORT} options can facilitate this, their improper use may violate TCP standards and introduce complex, hard-to-debug issues. Also, the repetitive SYN transmissions inherent in traditional TCP hole punching simultaneous open attempts can be misconstrued as SYN flood attacks by network security systems, leading to blocked connections or temporary IP blacklisting. TCP’s stateful nature can also consume router resources under repeated failed attempts, potentially degrading network performance, leading to denial-of-service conditions in constrained network environments or prohibiting subsequent retries. 

The findings of our empirical study on transport protocol dependence present a counterpoint to this conventional understanding. Our data show similar success rates, around $70\%$, for both TCP and QUIC. This contradicts the long-held common assumption that ``UDP hole punching is easier than TCP''. The reconciliation of this non-intuitive finding with the detailed theoretical difficulties of TCP hole punching suggests DCUtR's synchronization effectiveness. 
By having the initiator wait for $\frac{1}{2} \cdot \text{RTT}_{\text{relayed}}$ before dialing, the protocol assumes approximately symmetrical network delays so that both peers' SYN packets depart before either arrives at the opposite NAT. Under this assumption, unsolicited SYN packets would only reach the other peer's NAT if the RTT measurement error exceeds the direct one-way latency between the two NATs. Our RTT accuracy measurements support this: the standard deviation is less than half of the measured RTT in the large majority of cases. When path asymmetry is significant, however, the synchronization margin shrinks, which may partly explain the remaining $\sim30\%$ failure rate. The synchronization accuracy means the problem space for DCUtR largely becomes deterministic rather than stochastic. This insight motivates the optimizations described in Section~\ref{sec:optimizations}.

\pb{Limitations.} As noted in Section~\ref{sec:network-identification}, approximately $29\%$ of collected data points are excluded because prerequisite stages, such as establishing a relayed connection via Circuit~v2 or discovering public addresses via Identify, failed before a hole-punch would be attempted. Our reported $70\%$ success rates, therefore, characterize the performance of the DCUtR hole-punching stage, \textit{conditional} on earlier steps succeeding. 

The unconditional end-to-end connectivity metric for establishing a direct connection from scratch is necessarily lower once these prerequisite failure modes are taken into account. 
Furthermore, as discussed in Section~\ref{sec:dataset:limitations}, our dataset does not support a definitive classification of remote NAT types, preventing us from attributing individual failures to specific NAT mapping or filtering behaviors rendering this a promising avenue for future work.

This distinction underscores the multiple challenges that arise when enabling P2P connectivity in the wild.



\pb{Security Considerations.} DCUtR's permissionless design introduces several security trade-offs. During the handshake, both peers disclose their public IP addresses and port numbers via CONNECT messages routed through the relay. Because any peer including malicious actors can initiate this exchange, systematic collection of these addresses through repeated hole-punch probes could facilitate targeted scanning or exploitation of inadvertently exposed services. The relay itself represents a trust boundary. It can observe the addresses and connection patterns of both parties, enabling behavioral profiling over time. On the denial-of-service front, repeated failed TCP simultaneous-open attempts can be misidentified as SYN floods by stateful firewalls, risking IP denylisting, while resource-constrained NATs may exhaust their session tables under sustained probing.

\section{Optimizations \& Future Work}
\label{sec:optimizations}

The remaining $\sim30\%$ failure rate represents the next frontier for DCUtR improvement. Failures can arise from transient network conditions, misconfigurations, overly restrictive firewalls, or the prevalence of Symmetric NATs. We propose a multi-tiered roadmap to universal connectivity. The viability of this roadmap is underscored by the central finding of transport-agnosticism of DCUtR as a result of the effective RTT-based synchronization.

\subsection{Probabilistic Traversal via the Birthday Paradox}
\label{sec:optimization:symmetric-nat}

The $\sim30\%$ failure rate, despite the protocol's general efficacy (H1), is likely due, in part, to challenging NAT configurations, such as Endpoint-Dependent Mappings (EDM). To address this class of failures, which cannot be solved by simple synchronization, we propose exploring techniques based on the birthday paradox, a known method for traversing EDM NATs by creating a high probability of a port collision. Halkes et al.~\cite{Halkes2011-og} observed in 2011 that approximately $11\%$ of their monitored peer cohort were behind EDM NAT devices. As discussed in Section~\ref{sec:network-address-translation}, these devices make it impossible to predict the assigned port when establishing connections to other peers in the network, rendering the port information exchanged via the relay in the DCUtR protocol ineffective. 

To address this challenge, we can exploit the statistical properties of port allocation to create collisions within the $2^{16}$ port search space that would otherwise be computationally intractable. For example, if the peer behind the EDM NAT opens 256 ports simultaneously while the connecting peer probes 256 randomly selected ports (representing $0.4\%$ of the total search space), the probability of establishing a successful connection reaches $64\%$. When the search space coverage increases to $3.1\%$ (2,048 ports), the success probability rises to $99.9\%$. However, when both peers operate behind EDM NATs, the success probability drops significantly to $0.01\%$ when probing 2,048 ports on one side and 256 on the other~\cite{Anderson_2020}.
Based on the EDM NAT prevalence reported by Halkes et al.~\cite{Halkes2011-og}, we expect to encounter approximately $19.6\%$ mixed EDM/EIM peer combinations and $1.2\%$ combinations where both peers are behind EDM NATs. If birthday paradox techniques achieve a $64\%$ success rate for the $19.6\%$ of mixed connection attempts, the overall DCUtR success rate could improve $12.5\%$.
We assume conservative values for port probing and opening operations, as NAT devices maintain limited active session capacities and may interpret intensive port scanning as malicious behavior, potentially blocking the peer's IP address for extended periods. Without these constraints, success rates could be substantially higher.

Beyond NAT resource limits, opening a high volume of ports (e.g., 256--2,048) simultaneously also increases a node's attack surface: 
each open port represents a potential entry point for fingerprinting or exploitation by adversaries who may be aware of the technique. 
Aggressive port allocation patterns may trigger network-level intrusion detection systems. 
Deployments of birthday-paradox-based traversal must therefore balance the probability of successful traversal against the security exposure introduced by maintaining many concurrent open ports.

The data from Halkes et al.~\cite{Halkes2011-og} presents limitations due to its age (2011) and relatively small sample size (fewer than 2,000 peers). Two conflicting trends have emerged since their study: \one the expected reduction in EDM NAT deployment as vendors increasingly comply with the BEHAVE RFC~\cite{rfc4787}, and \two the proliferation of Carrier-Grade NAT (CGNAT) deployments, which employed EDM translation in $40\%$ of cases as of 2016~\cite{Richter2016}. Future empirical studies are necessary to establish more accurate contemporary estimates of expected success rate improvements.

\subsection{Protocol Optimizations}
\label{sec:optimization:protocol}
\pb{Refining RTT Calculation by Removing Local Latency.} A powerful optimization would let users choose a proxy with advantageous NAT properties (e.g., a full-cone NAT). This strategy, however, introduces an additional network hop, making the timing of the hole punch even more critical. The current DCUtR protocol utilizes Round Trip Time (RTT) measurements between \textbf{peers} to coordinate hole punching attempts. However, it is the RTT between the respective NAT devices, rather than the peers themselves, that is truly pertinent. Although our measurements indicate that peer-to-peer RTT often serves as a reasonable approximation, this approach may be insufficient in more complex network topologies. In such cases, the resulting discrepancy can cause the \texttt{SYNC} packet to arrive outside the optimal time window, thereby increasing the likelihood of connection failures. Therefore, we propose an enhancement where the listener includes its observed  $\text{RTT}_{\text{Listener - NAT}}$ to its NAT (or an equivalent near-edge network element) within its \texttt{CONNECT} message. The initiator, upon receiving this information, can then compute a more accurate wait time $T_{\text{wait}}$ for sending its \texttt{SYNC} packet by also incorporating the $\text{RTT}_{\text{Initiator - NAT}}$ to its NAT. The refined calculation for the protocol's RTT then reads as follows: $T_{\text{wait}} = 1/2( \text{RTT}_{\text{Listener - Initiator}} + \text{RTT}_{\text{Listener - NAT}} -\text{RTT}_{\text{Initiator - NAT}} )$.
This approach aims to better estimate the one-way traversal to the remote peer's NAT by factoring out local latency on each side, thus improving the timing precision of the \texttt{SYNC} message in topologies where the local network segment between a peer and its NAT introduces significant delay.

\pb{Alternating Roles on Connection Retries.} The validation of H4 reveals that retries exhibit diminishing returns. This suggests that repeating the same action is suboptimal. We propose an alternative retry strategy for subsequent attempts to increase the success rate. In the case of QUIC hole punching, upon receiving the SYNC message from the initiator, the listener immediately dials back. The initiator, on the other hand, starts to send UDP packets filled with random bytes to the listener upon expiry of the 1/2 RTT timer to prime its NAT. This will result in a QUIC connection where the listener is the client and the initiator is the server. If we assume that the initiator is behind an EDM NAT but the listener is not, the listener's client hello will not be able to traverse the initiator's NAT because of the unpredictable nature of the port mappings. We propose to alternate the roles in subsequent attempts. The above scenario would succeed on the second try, given that the initiator is not behind a symmetric NAT. Contrary to a TCP simultaneous open, having both parties sending client hellos might result in two distinct QUIC connections, which can also be acceptable in some scenarios.

\pb{Proactive NAT Priming to Prevent Denylisting.} Some NATs or firewalls denylist external IPs if inbound UDP packets arrive before any outbound state exists~\cite{rfc5128}.
To mitigate this, we propose a more cautious NAT priming strategy. Instead of sending random UDP payloads that might reach the initiator’s NAT, the listener should send low Time-To-Live (TTL) packets (e.g., TTL=3) to the initiator’s advertised addresses. This limits propagation while priming the listener's NAT. Similarly, once the initiator receives the listener’s \texttt{CONNECT} message, it should send low-TTL packets to the listener until its \texttt{SYNC} timer fires. This staggered priming reduces denylisting risk while establishing outbound mappings before hole punching begins.

\subsection{Ecosystem Strategies}
\label{sec:optimization:ecosystem}



Long-term success also depends on improving the network environment itself through a combination of standards advocacy and better use of existing mechanisms. This echoes a long-standing hope in the community (e.g. by Halkes et al.~\cite{Halkes2011-og} in 2011), that broader vendor compliance with standards like IETF's BEHAVE~\cite{rfc4787} would reduce the prevalence of such ``hard'' NATs. In parallel, our findings confirm that Connection Reversal optimization is highly effective when a port mapping is available (H4). Client software should actively assist users in enabling UPnP/PMP on their routers, explaining the direct connectivity benefits while acknowledging potential security trade-offs.
\section{Related Work}
\label{sec:related-work}

NAT hole punching, or NAT traversal, has been a recognized technique for enabling direct peer-to-peer communication across Network Address Translators for over two decades. The concept was initially introduced by Dan Kegel in 1999~\cite{kegel_1999}, who outlined a NAT traversal protocol utilizing UDP packets, primarily motivated by the requirements of peer-to-peer gaming applications. One of the foundational publications in this domain is by Ford et al.~\cite{Ford2005PeertoPeerCA}, which describes early approaches to peer-to-peer communication across NATs. Subsequent research has explored various facets of NAT traversal. Guha and Francis~\cite{Guha2005} provided an empirical characterization and measurement of TCP traversal through NATs and firewalls, offering insights into the challenges posed by TCP's connection-oriented nature. Similarly, Halkes and Pouwelse~\cite{Halkes2011-og} investigated UDP NAT and firewall puncturing in real-world scenarios, contributing to the understanding of UDP's behavior in NAT environments. The Interactive Connectivity Establishment (ICE) protocol~\cite{rfc8445, rfc5245} unifies several NAT traversal techniques, including Session Traversal Utilities for NAT (STUN)~\cite{rfc8489} and Traversal Using Relays around NAT (TURN)~\cite{rfc8656}, to provide a comprehensive framework for establishing peer-to-peer connections in complex network topologies. ICE is widely adopted in real-time communication protocols, such as WebRTC~\cite{rfc8825}.

Further studies have delved into the specific aspects and challenges of NAT traversal. Ford et al.~\cite{Ford2005PeertoPeerCA} provided a comprehensive overview of the state of peer-to-peer communication across NATs, detailing various techniques and their limitations. Richter et al.~\cite{Richter2016} conducted a multi-perspective analysis of Carrier-Grade NAT (CGNAT) deployment, highlighting its prevalence and impact on network behavior. Other works have explored edge-case integration into established NAT traversal techniques~\cite{Keller2022}, knowledge-based NAT-traversal for home networks~\cite{muller2008}, and the revisiting of NAT hole punching strategies~\cite{Maier2011}. Additionally, research has examined peer NAT proxies for peer-to-peer games~\cite{Seah2009}. Liang et al.~\cite{liang2024} show that QUIC-based hole punching is theoretically 0.5 RTT faster than TCP-based hole punching due to its integrated 1-RTT handshake compared to TCP's 3-way handshake followed by a separate TLS handshake.
\section{Conclusion}
\label{sec:conclusion}
This paper presents a large-scale empirical evaluation of DCUtR, validating its design principles and establishing a contemporary benchmark for decentralized NAT traversal. 

We show that hole punching succeeds at a conditional rate of approximately $70\%$, given successful relay reservation and address discovery (H1), that this performance is independent of relay characteristics (H2), and that synchronization is highly effective, with most successes occurring on the first attempt (H3a). Contrary to long-standing belief, TCP and QUIC exhibit comparable success rates (H3b). We also validate key protocol optimizations such as Connection Reversal (H4) and demonstrate significant latency reductions after successful traversal.Our CC BY-SA licensed dataset of over 4.4 million measurements across more than 85k networks globally provides a valuable resource for future research.

DCUtR meaningfully advances decentralization by reducing reliance on centralized intermediaries, thus mitigating single points of failure and enhancing self-sovereign control over data. While not a full replacement for STUN/TURN today, it can already offload significant traffic and, when combined with suitable relays (e.g. VPN or SSH gateways), has the potential to eliminate dependence on these services entirely. The proposed optimizations of role alternation on retries, refined RTT calculation, proactive NAT priming, and the birthday paradox exploitation offer concrete paths toward higher direct-connectivity rates and strengthen the protocol as a building block for the decentralized Internet.

\begin{acks}
We thank Max Inden for his invaluable contributions in proposing the honeypot component, building the Rust client, and suggesting the alternating roles on connection retries optimization. We also extend our sincere thanks to Elena Frank for her dedicated work on the Rust client. 
\end{acks}

\clearpage
\bibliographystyle{ACM-Reference-Format}
\bibliography{references}

\appendix
\clearpage
\onecolumn
\section{Appendix}

\subsection{DCUtR Sequence Diagram}
\label{appendix:dcutr-sequence-diagram}

\begin{figure}[h]
    \centering
    \includegraphics[width=\textwidth]{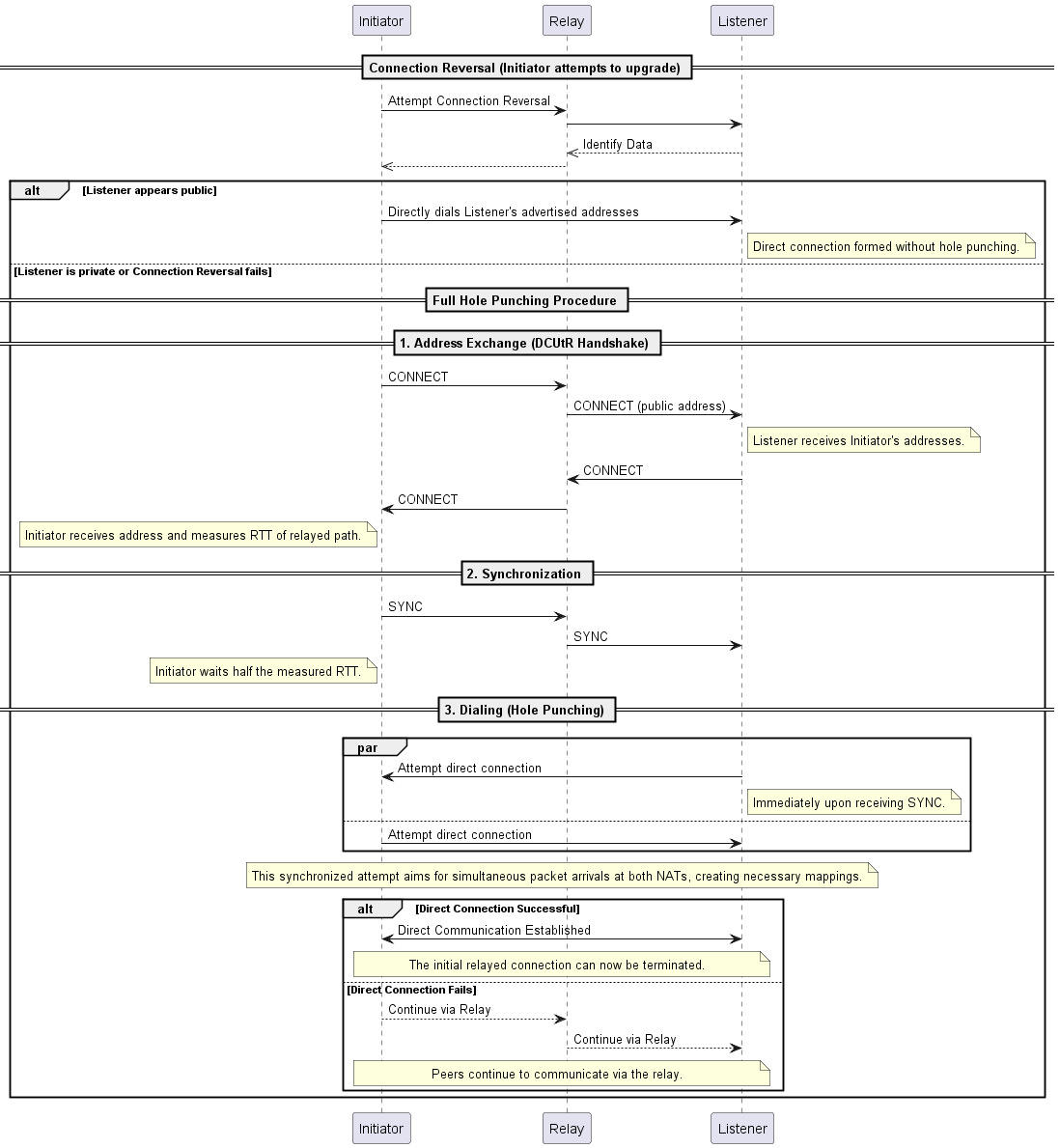}
    \caption{DCUtR Protocol Sequence Diagram}
    \label{fig:dcutr-sequence-diagram}
\end{figure}
\subsection{Dataset}
\label{appendix:dataset}

In this section, we will provide a detailed overview of the dataset used in our analysis. The provided dataset is a Postgres database dump available via this IPFS CID

\begin{center}
    \href{https://bafybeia7sq3nfd7c4obcy7ahjvnoka7ujdiob33r7rqyeycgicdt3iknki.ipfs.dweb.link/}{\texttt{bafybeia7sq3nfd7c4obcy7ahjvnoka7ujdiob33r7rqyeycgicdt3iknki}}
    \vspace{0.1cm}
\end{center}

\begin{figure}
    \centering
    \includegraphics[width=\textwidth]{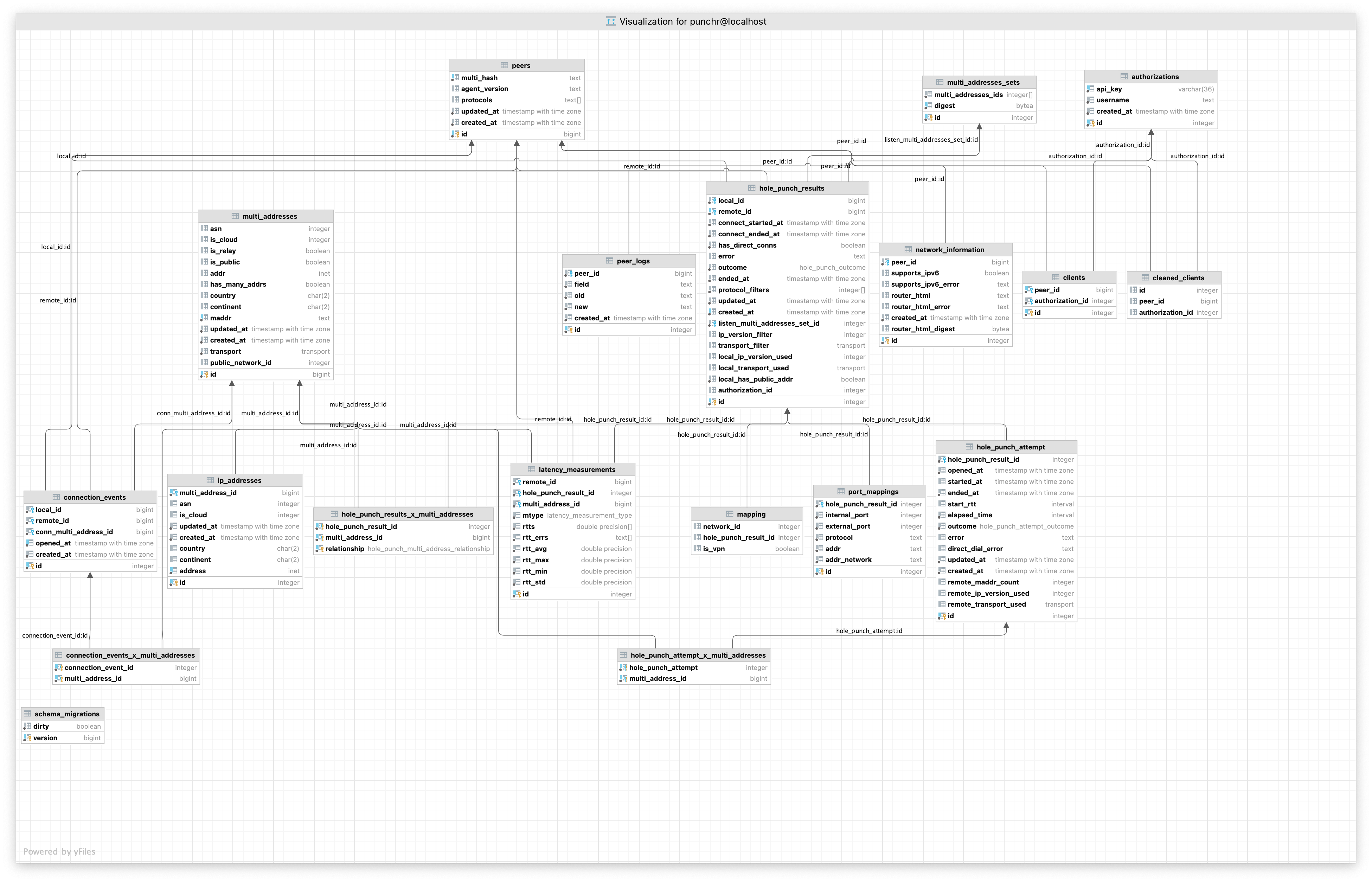}
    \caption{Postgres database UML diagram}
    \label{fig:psql-db-uml}
\end{figure}

Figure~\ref{fig:psql-db-uml} shows the corresponding UML diagram. This includes information on the time scope of the data, as well as instructions on how to download the dataset and set up a database for storage. Additionally, we will describe the process for restoring the data, and how we obtained the database dump.

\subsubsection{Peers/Multiaddresses}

We collect agent version, supported protocols, and Multihashes of all peers that interact with our measurement infrastructure. Similarly, when we get hand of a multi address we extract the underlying IP address, corresponding Geolocation (continent, country), if it’s a public address or not, if it’s a relayed address or not, if the IP address belongs to a known datacenter, and the corresponding Autonomous System Number.

\subsubsection{Connection Events}

As discussed in Section~\ref{sec:measurement}, the honeypot component tracks inbound connections from DCUtR capable peers. We call these events ``Connection Events''. A Connection event consists of the following information:

\begin{itemize}
    \item \texttt{local\_id} -- The internal database peer ID of the honeypot
    \item \texttt{remote\_id} -- The internal database peer ID of the peer that connected to the honeypot
    \item \texttt{conn\_multi\_address\_id} -- The multiaddress of the connection to the honeypot
    \item \texttt{connection\_events\_x\_multi\_addresses} -- All advertised multi addresses of the remote peer. They strictly won't contain a public IP address that is not a relayed address. This set will be served to clients.
\end{itemize}

\subsubsection{Hole Punch Results/Attempts}

A ``Hole Punch Result'' is what gets reported back by the clients and can consist of multiple ``Hole Punch Attempts''. Both data points can have different outcomes that are listed in Table~\ref{tab:hpr-outcomes} and Table~\ref{tab:hpa-outcomes}.

\subsubsection{Latency Measurements}

Before clients attempt to hole punch a remote peer, the client measures several latencies. It measures the latencies to all relays that the remote peer claims to be reachable through. It also measures the latency to the remote peer through one of the relays. Finally, if the hole punch succeeds, the client also measures latency of the direct connection. The database field \texttt{mtype} is \texttt{TO\_RELAY}, \texttt{TO\_REMOTE\_THROUGH\_RELAY}, and \texttt{TO\_REMOTE\_AFTER\_HOLEPUNCH} respectively.

\subsubsection{Port Mappings}

For each hole punch result, we track any active port mappings that get reported back from AutoNAT. If a port mapping is in place it is more likely for an outcome of \texttt{CONNECTION\_REVERSED}. With this data we can test this hypothesis. Each port mapping consists of the following data:

\begin{itemize}
    \item The hole punch it refers to
    \item Internal Port of the client in the local network
    \item External Port of the router
    \item Transport used for port mapping
    \item External facing address
\end{itemize}

\subsubsection{Authorizations/Clients}

Each time someone signed up through our Google Form we generated a UUID API-Key and saved it alongside the email address (from the form) into this table. We asked participants to provide the API Key to their client installations (GUI or CLI). Every hole punch result reported from these clients contained the API key so that we can associate the result with the Google Form information.

Importantly, the Go-Client generated ten peer identities upon startup. Each peer ID would listen on a different port. For each hole punch we ‘round robin’ through the ten peers. This way we mitigated persistent port mappings to yield a high number of \texttt{CONNECTION\_REVERSED} results. When the client starts up it reports the identities of these ten clients to the server. This fills the `clients` table where you have a mapping between peer ID and authorization ID.

So, to map a hole punch result to an authorization, you’d need to map the \texttt{local\_id} field onto the \texttt{clients} table and then to the \texttt{authorizations} table.



\subsection{Ethics}
\label{sec:ethics}

The research followed strict ethical guidelines. Participants either signed up, receiving detailed information and providing explicit consent regarding data collection (including IP addresses and libp2p PeerIDs) and its academic use, or voluntarily ran a tool whose documentation (README) clearly stated data collection (including IP addresses and libp2p PeerIDs) for research.

Initially, IP addresses were stored in clear text for network analysis (e.g., geographic distribution, network identification for the DCUtR protocol) and securely on a restricted-access research server, accessible only to the core researcher. After core analysis for this paper, all IP addresses were irreversibly anonymized or deleted, preventing deanonymization while allowing statistical correlation.

Data from remote, non-consenting peers was collected based on their operation of public, permissionless P2P nodes (IPFS, libp2p), which inherently expose network metadata (IP, PeerID, protocols) through standard interactions. The ``honeypot'' acted as a regular libp2p node, observing public addresses and sending standard protocol messages. Collected data (e.g., hole punch success) was a byproduct of these standard interactions, with no non-standard or privacy-invasive data gathered beyond what a public P2P node inherently exposes.

\end{document}